\documentclass[a4paper]{article}
\pdfoutput=1
\usepackage[bookmarks=false]{hyperref}
\usepackage{hyperref}
\usepackage[utf8]{inputenc}

\usepackage{tikz}
\usepackage{graphicx}
\usepackage{bm}
\usepackage{amsmath}
\usepackage{fullpage}

\begin{document}

\title{Could quantum decoherence and measurement be deterministic phenomena?}

\author{Jean-Marc Sparenberg\footnote{Corresponding author:
\href{mailto:jmspar@ulb.ac.be}{jmspar@ulb.ac.be}},
Réda Nour
and Aylin Manço
\\
Université libre de Bruxelles (ULB),
Nuclear Physics and Quantum Physics, CP 229, \\
École polytechnique de Bruxelles, B 1050 Brussels, Belgium}

\maketitle

\begin{abstract}
The apparent random outcome of a quantum measurement is conjectured
to be fundamentally determined by the microscopic state
of the macroscopic measurement apparatus.
The apparatus state thus plays the role of a hidden variable which,
in contrast with variables characterizing the measured microscopic system,
is shown to lead to a violation of Bell's inequalities
and to agree with standard quantum mechanics.
An explicit realization of this interpretation is proposed
for a primitive model of measurement apparatus inspired by Mott \cite{mott:1929}:
in the case of an $\alpha$-particle spherical-wave detection in a cloud chamber,
the direction of the observed linear track
is conjectured to be determined by the position of the atoms of the gas filling the chamber.
Using a stationary-state coupled-channel Born expansion,
a reduction of the spherical wave function is shown to be necessary to compensate the flux loss
due to scattering on the chamber atoms.
Being highly non local, this interpretation of quantum mechanics is finally argued
to open the way to faster-than-light information transfer.
\end{abstract}

\section{Introduction: advocacy for determinism}

The most disturbing feature of quantum physics is probably the fundamental randomness
of the outcome of a quantum measurement:
the repeated measurement of the same observable
on a microscopic system prepared in the same state,
with the same apparatus,
can lead to very different results.
Think for instance about the detection of an $s$-wave $\alpha$-particle in a cloud chamber.
Though the wave function is perfectly spherical [see equation \eqref{sph} below],
the observed individual tracks are linear,
spread isotropically in apparently totally random directions.

This seemingly unavoidable lack of determinism very deeply bothered Einstein himself,
as expressed by his famous opinion, expressed in a 1926 letter to Born,
that ``[God] does not throw dice'' \cite{born:1971}.
Indeed, determinism is one of the keystones of modern science,
as experiments are only reproducible provided the same cause always produces the same effect.
Giving up this very principle is in a sense equivalent to giving up science itself!
Another motivation for not giving up determinism is timeless theories.
In these, time does not have a fundamental character but rather emerges as a secondary quantity
(see in particular the contribution by Barbour at this conference and reference \cite{barbour:01}).
In a deterministic world, the present moment is unambiguously related to the past
from which it results and to the future which it determines,
hence strengthening the idea that past, present and future are one and the same.

A possible way to recover determinism in quantum mechanics is to assume,
in the spirit of Einstein, Podolsky and Rosen \cite{einstein:1935},
that the knowledge of a microscopic system provided by quantum mechanics is incomplete.
Hidden variables should exist, that determine the measurement outcome,
even though they might not be accessible to the experimenter.
Another fundamental advantage of hidden variables in Einstein's view
is that they avoid violation of special relativity by quantum mechanics:
instantaneous correlations between distant particles in an intricated state
can be explained by a common variable shared by the particles when they first interacted.

The very logical assumption of hidden variables was however proved incorrect.
Bell \cite{bell:1964} (also reproduced in \cite{bell:01})
and others \cite{bohm:1957,bell:1966,clauser:1969}
established inequalities that should be satisfied by measurement outcomes on intricated particles
if such hidden variables existed,
and that are violated by quantum mechanics predictions.
Aspect and others \cite{freedman:72,aspect:81,aspect:82} then checked experimentally
that quantum predictions were correct,
which ruled out hidden variables at least in their simplest form.
These experiments also prove that quantum mechanics is highly non local
and that the measurement performed on one particle indeed has an instantaneous impact
on the measurement performed on another particle intricated with the first one,
even at very large distances.
The lack of determinism of quantum measurements then comes as a ``savior'' of special relativity \cite{eberhard:78}:
since the outcome of the measurement on the first particle is random,
the intrication between both particles, though it exists and is instantaneous,
cannot be used to transmit information faster than light.

In the present work, we try to answer the question:
what actually determines the result of a given quantum measurement?
In the case of the $\alpha$-particle detection,
what actually determines the direction of an observed track?
We explore the hypothesis that hidden variables are not variables characterizing the microscopic
system under study (the $\alpha$-particle in the cloud chamber experiment
or the pair of intricated particles in an EPR-type experiment),
but rather the macroscopic apparatus (or apparatuses) used to study them.
Indeed, the mysterious properties of quantum mechanics (randomness and non locality)
always occur when a microscopic system, well described by its wave function,
interacts with a highly complex and unstable macroscopic system.
Describing the state of such a complex system at the microscopic level is of course very difficult
as its internal degrees of freedom (or more generally its ``environment'') cannot be monitored
and appear as totally random to our ``macroscopic eyes''.
It is however tempting to assume, as also suggested by Gaspard \cite{gaspard:01},
that it is precisely this microscopic state of the system
which is at the origin of the randomness observed during the measurement process.

Our approach is also related to attempts to describe quantum measurement processes based on decoherence
\cite{zurek:81,zurek:91,zurek:02}.
There also, it is the interaction with a macroscopic system,
e.g.\ a measurement apparatus,
that makes a microscopic system initially in a linear superposition of states
evolve into a standard statistical mixture of possible outcomes.
Mathematically, decoherence is usually described by {\em ad hoc} terms in a master equation,
with no attempt to explain particular outcomes of the decoherence process into one or another result.
In the present approach, we go a step further, similarly to Gaspard and Nagaoka \cite{gaspard:99}:
the internal state of the macroscopic system precisely determines {\em which} outcome will occur,
not only the mere fact that probabilistically-acceptable outcomes do occur.

The aim of the present work is to test whether this simple deterministic interpretation leads to fundamental
principle impossibilities or not.
In the following, we first show that it leads to violations of Bell's inequalities,
in agreement with standard quantum mechanics and with experimental results.
The price to pay for that success is a strong non locality,
which is not saved any more by quantum randomness \cite{corbett:08}.
Next we come back to the problem of $\alpha$-ray tracks in a cloud chamber.
We explore the hypothesis that the positions of the atoms of the chamber
actually determine the direction of the observed track,
hence playing the role of apparatus hidden variables
randomly distributed because of the thermal agitation inside the chamber.
Possible experimental tests of this idea are then proposed.
Finally,
this interpretation is speculated, if correct, to open the way to faster-than-light communications.

\section{Apparatus hidden variables and Bell's inequalities}

Let us first recall the principle of the Einstein-Podolsky-Rosen thought experiment \cite{einstein:1935},
as formulated by Bohm-Aharonov \cite{bohm:1957}.
A system of two particles, 1 and 2, both with spin $\frac{1}{2}$ denoted as $\bm{s}_1$ and $\bm{s}_2$,
is created in the maximally entangled singlet (i.e.\ spin 0) state
\begin{equation} \label{singlet}
|0 0\rangle
= \frac{1}{\sqrt{2}} \big[ |+\rangle_{\bm{\hat{u}}1} |-\rangle_{\bm{\hat{u}}2}
- |-\rangle_{\bm{\hat{u}}1} |+\rangle_{\bm{\hat{u}}2} \big], \qquad \forall \bm{\hat{u}}, 
\end{equation}
where $\bm{\hat{u}}$ is an arbitrary unit vector.
The particles are assumed to stay in that spin state while moving away from each other.
The spin of particle 1 is then measured in direction $\bm{\hat{a}}$
while the spin of particle 2 is measured in direction $\bm{\hat{b}}$,
each measurement leading to the result $+\frac{\hbar}{2}$ or $-\frac{\hbar}{2}$.

Following Bell \cite{bell:1964}, we define the mean correlation between spin 1 and 2
measured along directions $\bm{\hat{a}}$ and $\bm{\hat{b}}$ as
\begin{equation}
E(\bm{\hat{a}}, \bm{\hat{b}}) \equiv
\tfrac{4}{\hbar^2} \big\langle (\bm{s}_1 \cdot \bm{\hat{a}}) (\bm{s}_2 \cdot \bm{\hat{b}})\big\rangle.
\end{equation}
For the singlet state \eqref{singlet}, this quantity can be shown to take the value
\begin{equation} \label{correl}
E_Q(\bm{\hat{a}}, \bm{\hat{b}}) = -\bm{\hat{a}} \cdot \bm{\hat{b}}
\end{equation}
in standard quantum mechanics.
In particular, when $\bm{\hat{a}}=\bm{\hat{b}}$, the correlation between both measurements is perfect
and one has $E_Q(\bm{\hat{a}}, \bm{\hat{a}}) = -1$,
whatever the distance between both particles at the time their spins are measured.
This non-local correlation, typical of entangled quantum states,
chagrined Einstein, hence leading to the hypothesis that the state of the particle pair
was given not only by the wave function \eqref{singlet} but also by some hidden variable $\lambda$.
This variable is assumed to be randomly distributed;
its value is supposed to determine the result of any of the measurements.
It thus also determines the value of correlation \eqref{correl},
which we now denote as $E_\lambda(\bm{\hat{a}}, \bm{\hat{b}})$.
Such a variable avoids ``spooky actions at a distance'' apparently implied by state \eqref{singlet}.
However, Bell \cite{bell:1964} brilliantly proved
that the very existence of variable $\lambda$ leads to inequalities for the correlations
calculated for three arbitrary directions $\bm{\hat{a}}, \bm{\hat{b}}, \bm{\hat{c}}$,
namely
\begin{equation} \label{bell}
|E_\lambda(\bm{\hat{a}}, \bm{\hat{b}}) - E_\lambda (\bm{\hat{a}},\bm{\hat{c}})|
\le 1 + E_\lambda(\bm{\hat{b}}, \bm{\hat{c}}).
\end{equation}
The observation by Bell is that these inequalities are violated by the quantum prediction
\eqref{correl} for particular directions $\bm{\hat{a}}, \bm{\hat{b}}, \bm{\hat{c}}$
(for instance differing by $\frac{\pi}{8}$),
hence offering the possibility
to discriminate between standard quantum mechanics and hidden variables experimentally.
Such experiments were realized with photon polarizations by Aspect {\it et al.}~\cite{aspect:82}
and displayed a clear disagreement with inequality \eqref{bell}
while confirming the quantum prediction \eqref{correl}.

This result somehow signs the death warrant of hidden variables characterizing the microscopic system
(in this case the pair of particles).
Let us now show that in contrast it does not prevent
the hypothesis of hidden variables characterizing the internal state of the measurement apparatus.
In the present situation, we introduce variables $\Lambda_1$ and $\Lambda_2$
that correspond to the apparatus measuring particle 1 and 2 respectively.
In the same spirit as Bell, we do not attempt to find the explicit physical nature of these variables;
we only assume that they exist and that they determine the experimentally observed results,
together with the particle state and the apparatus orientation.
Let us denote by
\begin{equation} \label{R}
 R(|\sigma\rangle, \bm{\hat{a}}, \Lambda) = \tfrac{2}{\hbar} \bm{s} \cdot \bm{\hat{a}}
\end{equation}
the result of the measurement of the projection of the spin $\bm{s}$ of one of the spin $\frac{1}{2}$ particles
assumed to be in state $|\sigma\rangle$,
in one of the measurement apparatus oriented along direction $\bm{\hat{a}}$.
This result is assumed to be determined by the value of the hidden variable $\Lambda$
characterizing the internal state of the apparatus.
The precise value of $\Lambda$ is not known but its probability distribution $p(\Lambda)$ has to satisfy
$\sum_\Lambda p(\Lambda)=1$,
where the sum covers all the possible values of $\Lambda$.
The two apparatuses are assumed to be identical;
a unique function \eqref{R} hence describes both of them.

This function has to satisfy the following conditions to describe the measurement process correctly:
\begin{itemize}
 \item it should only take the values $\pm 1$ as the particle has spin $\frac{1}{2}$;
 \item the measurement of a polarized state made along its polarization direction should be reproducible, 
whatever the internal state of the apparatus;
hence
\begin{equation}
R(|\pm\rangle_{\bm{\hat{a}}}, \bm{\hat{a}}, \Lambda) = \pm 1, \qquad \forall \Lambda;
\end{equation}
 \item the mean value of a state polarized along direction $\bm{\hat{a}}$,
as measured through an apparatus oriented along direction $\bm{\hat{b}}$,
should be related to the angle between $\bm{\hat{a}}$ and $\bm{\hat{b}}$ as
\begin{equation}\label{Lamcon}
 \sum_\Lambda p(\Lambda) R(|+\rangle_{\bm{\hat{a}}}, \bm{\hat{b}}, \Lambda) = \bm{\hat{a}} \cdot \bm{\hat{b}}.
\end{equation}
\end{itemize}
The last two conditions are compatible with each other and condition \eqref{Lamcon} also implies that
\begin{equation}\label{Lamconbis}
 \sum_\Lambda p(\Lambda) R(|-\rangle_{\bm{\hat{a}}}, \bm{\hat{b}}, \Lambda) = -\bm{\hat{a}} \cdot \bm{\hat{b}}.
\end{equation}

Within these hypotheses, the mean correlation between the two measurements reads
\begin{equation} \label{correlLam}
E_{\Lambda_1 \Lambda_2}(\bm{\hat{a}}, \bm{\hat{b}})
= \textstyle \sum_{\Lambda_1 \Lambda_2} p(\Lambda_1) p(\Lambda_2)
R\big(|\sigma_1\rangle, \bm{\hat{a}}, \Lambda_1\big)
R\big(|\sigma_2\rangle, \bm{\hat{b}}, \Lambda_2\big),
\end{equation}
where $|\sigma_i\rangle$ is the state of particle $i$ as measured by apparatus $i$.
To fix ideas, we assume that the measurement in apparatus 1, oriented in direction $\bm{\hat{a}}$,
is made before that in apparatus 2, oriented in direction $\bm{\hat{b}}$
(but the time interval between both measurements can be made arbitrarily small).
Hence we choose to write the singlet state \eqref{singlet} with $\bm{\hat{u}}=\bm{\hat{a}}$.
The state of particle 1, as seen by apparatus 1, is thus
$|\sigma_1\rangle = \frac{1}{\sqrt{2}}
[|+\rangle_{\bm{\hat{a}}} - |-\rangle_{\bm{\hat{a}}}]$,
which implies that the first measurement gives $+1$ or $-1$ in 50\% of the cases.
This allows an easy calculation of the first sum in \eqref{correlLam}.
Immediately after this first measurement,
particle 2 is put in either a down or an up state along direction $\bm{\hat{a}}$,
depending on the result obtained for the first measurement.
The second apparatus being oriented along direction $\bm{\hat{b}}$,
the second sum in \eqref{correlLam} has thus to be calculated using equations \eqref{Lamcon} and \eqref{Lamconbis}.
One has finally
\begin{equation}
E_{\Lambda_1 \Lambda_2}(\bm{\hat{a}}, \bm{\hat{b}}) 
= \textstyle \frac{1}{2} \sum_{\Lambda_2} p(\Lambda_2) \left[R\big(|-\rangle_{\bm{\hat{a}}}, \bm{\hat{b}}, \Lambda_2\big)
- R\big(|+\rangle_{\bm{\hat{a}}}, \bm{\hat{b}}, \Lambda_2\big)\right]
= \textstyle \frac{1}{2} (-\bm{\hat{a}} \cdot \bm{\hat{b}}- \bm{\hat{a}} \cdot \bm{\hat{b}})
= -\bm{\hat{a}} \cdot \bm{\hat{b}}.
\end{equation}
The result is thus identical to the standard quantum mechanics prediction \eqref{correl}.
Apparatus hidden variables hence also lead to a violation of Bell's inequalities \eqref{bell},
in agreement with experiment.

While still agreeing with standard quantum mechanics,
the present theory is thus deterministic,
as the measurement results are determined by the actual values of $\Lambda_1$ and $\Lambda_2$.
Another advantage is that these hidden variables are local,
in the sense that $\Lambda_1$ only characterizes the state of apparatus $1$
and $\Lambda_2$ only characterizes the state of apparatus $2$.
Hence
there is no need to consider generalized inequalities for non-local hidden variables \cite{leggett:03}.
The price to pay for this simplicity is in contrast a strong non-locality.
The measurement performed in apparatus 1 is actually a measurement of the particle {\em pair} as a whole,
not only of particle 1: it immediately affects the state of both particles by reducing their wave function
to either $|+\rangle_{\bm{\hat{a}}1} |-\rangle_{\bm{\hat{a}}2}$
or $|-\rangle_{\bm{\hat{a}}1} |+\rangle_{\bm{\hat{a}}2}$.
The very structure of wave function \eqref{singlet} thus allows an actual ``spooky action at a distance''
which implies the perfect correlations between the states of both particles.

The consequences of this irreducible quantum non locality will be discussed in a relativistic perspective 
in the speculative conclusions.
Before that, we would like to consider a model of measurement apparatus
in which the physical nature of these internal degrees of freedom can be explicitly described,
to test the hypothesis that they determine the measurement outcome.
To do that, we consider what is probably the simplest possible measurement apparatus to model schematically,
namely the cloud chamber mentioned in the introduction.
There the $\Lambda$ internal variables could just be the positions of the atoms in the chamber.
Note that a schematic model of Stern-Gerlach apparatus could be built on the same basis,
with the screen being replaced by a cloud chamber.
As for photon polarization measurement apparatuses, a microscopic model should be based
on internal electronic degrees of freedom;
mesoscopic or cold polarisers might be interesting to consider in that respect.

\section{A schematic model for deterministic quantum measurement}

In the present section, we revisit an idea proposed by Mott \cite{mott:1929}
(also reproduced in \cite{mott:95}) and discussed in \cite{bell:81} (also reproduced in \cite{bell:01}),
namely the measurement of a spherical-wave $\alpha$-particle state in a Wilson cloud chamber.
Mott shows that the observation of linear tracks can be explained by the interaction of the
$\alpha$-particle spherical wave with the atoms of the gas filling the chamber.
Here, we reproduce his calculation but interpret his result differently,
hence leading to the interpretation
that the presence of the gas might not only explain the appearance of linear tracks
but also determine {\em which} linear track is actually observed
for a given microscopic configuration of the gas atoms.
We thus claim that Mott's model might be considered as the first model
for a deterministic decoherent process of wave-function reduction appearing in a quantum measurement.

\begin{figure}
\setlength{\unitlength}{6mm}
\begin{center}
(a)
\parbox{5cm}
{\begin{picture}(8,7)(-3,-0.5)
 \put(0,0){\circle*{0.7}}
 \put(3,1.5){\circle{2}}
 \put(3,6){\circle*{0.4}}
 \put(0,0){\vector(1,2){2.92}}
 \put(0,0){\vector(2,1){3}}
 \put(3,1.5){\vector(0,1){0.8}}
 \put(-3,-0.3){$\alpha$ emitter}
 \put(3.5,5.8){$\alpha$}
 \put(3,2.7){atom}
 \put(0.5,3){$\bm{R}$}
 \put(1.2,0){$\bm{a}$}
 \put(3.2,1.6){$\bm{r}$}
\end{picture}}
\qquad \qquad \qquad
(b)
\parbox{5cm}
{\begin{picture}(5.5,7)(-0.5,-0.5)
 \put(0,0){\circle*{0.7}}
 \put(4,4){\circle{0.6}}
 \put(3,1.5){\circle{0.6}}
 \put(3,6){\circle*{0.4}}
 \put(0,0){\vector(1,2){2.92}}
 \put(0,0){\vector(1,1){4.1}}
 \put(0,0){\vector(2,1){3.1}}
 \put(0.5,3){$\bm{R}$}
 \put(1.8,0.4){\mbox{$\bm{a}$}}
 \put(3.5,2.8){\mbox{$\bm{b}$}}
\end{picture}}
\end{center}
\caption{\label{fig:syst} Theoretical systems used to describe the emission of an $\alpha$ particle
by an $\alpha$ emitter in a cloud chamber.
Nuclei are represented as filled circles;
cloud-chamber atoms are represented as empty circles.
Only (a) the atom (b) the two atoms nearest to the $\alpha$ emitter are considered.}
\end{figure}
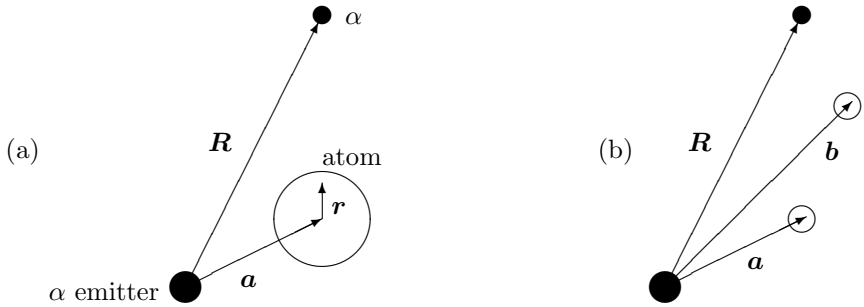

Let us first consider the interaction between an $\alpha$ $s$-wave,
emitted by a typical spherical $\alpha$ emitter like $^{210}$Po,
and a single atom in a cloud chamber [see figure \ref{fig:syst}(a)].
We assume that this atom, which is the first obstacle seen by the $\alpha$ wave,
is immobile at a fixed position $\bm{a}$.
This is justified as the typical thermal agitation energy of molecules in a cloud chamber (a few meV)
is much lower than the typical $\alpha$-particle kinetic energy $E_\alpha$ (a few MeV).
This obstacle is rather generic; its precise nature will not affect the general conclusions
drawn below.
To simplify calculations, we assume that it is described by an internal Hamiltonian $H$
with internal variables $\bm{r}$ and we only consider two excitations levels:
its ground state $E_0$ and its first excited state $E_1$.
The corresponding wave functions $\psi_0(\bm{r})$ and $\psi_1(\bm{r})$ are orthonormal,
i.e.\ $\langle \psi_i | \psi_j \rangle = \delta_{ij}$ for $i,j=0,1$;
they are assumed to be localised in a region of size $s$,
e.g.\ the typical size of an atom.
The total hamiltonian of the system is thus
\begin{equation}
 H = T_\alpha + H + V(\bm{R},\bm{r}),
\end{equation}
where $T_\alpha$ is the kinetic energy of the $\alpha$ particle
and $V(\bm{R},\bm{r})$ describes the interaction between the $\alpha$ particle and the obstacle.
In the following, this interaction is assumed to be small,
which allows the use of a Born-expansion perturbative treatment \cite{taylor:72}.

\begin{figure}
\setlength{\unitlength}{1cm}
\centerline{\begin{picture}(12,6)
 \put(-1,0){\scalebox{0.38}{\includegraphics{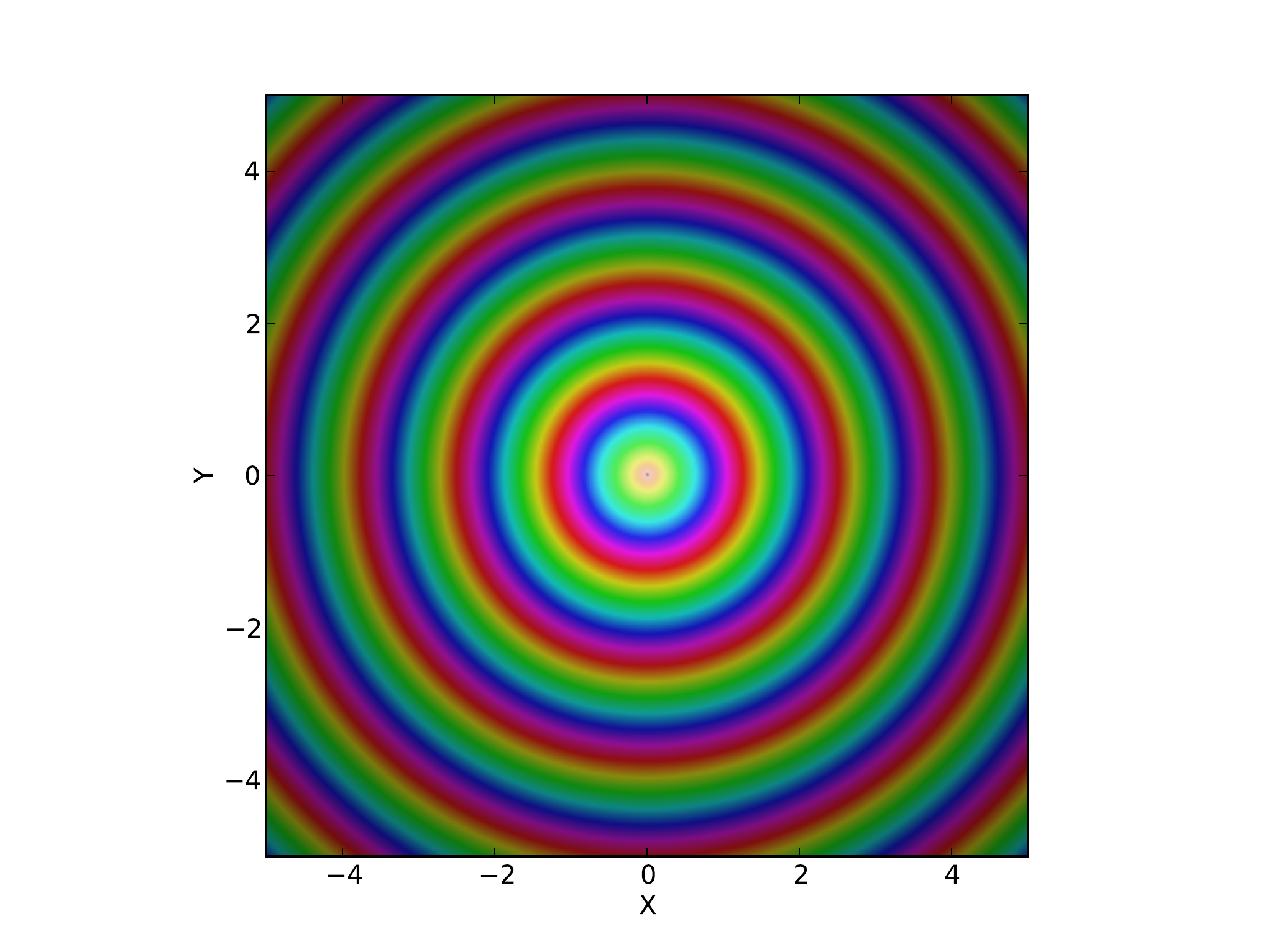}}}
 \put(5.5,0){\scalebox{0.38}{\includegraphics{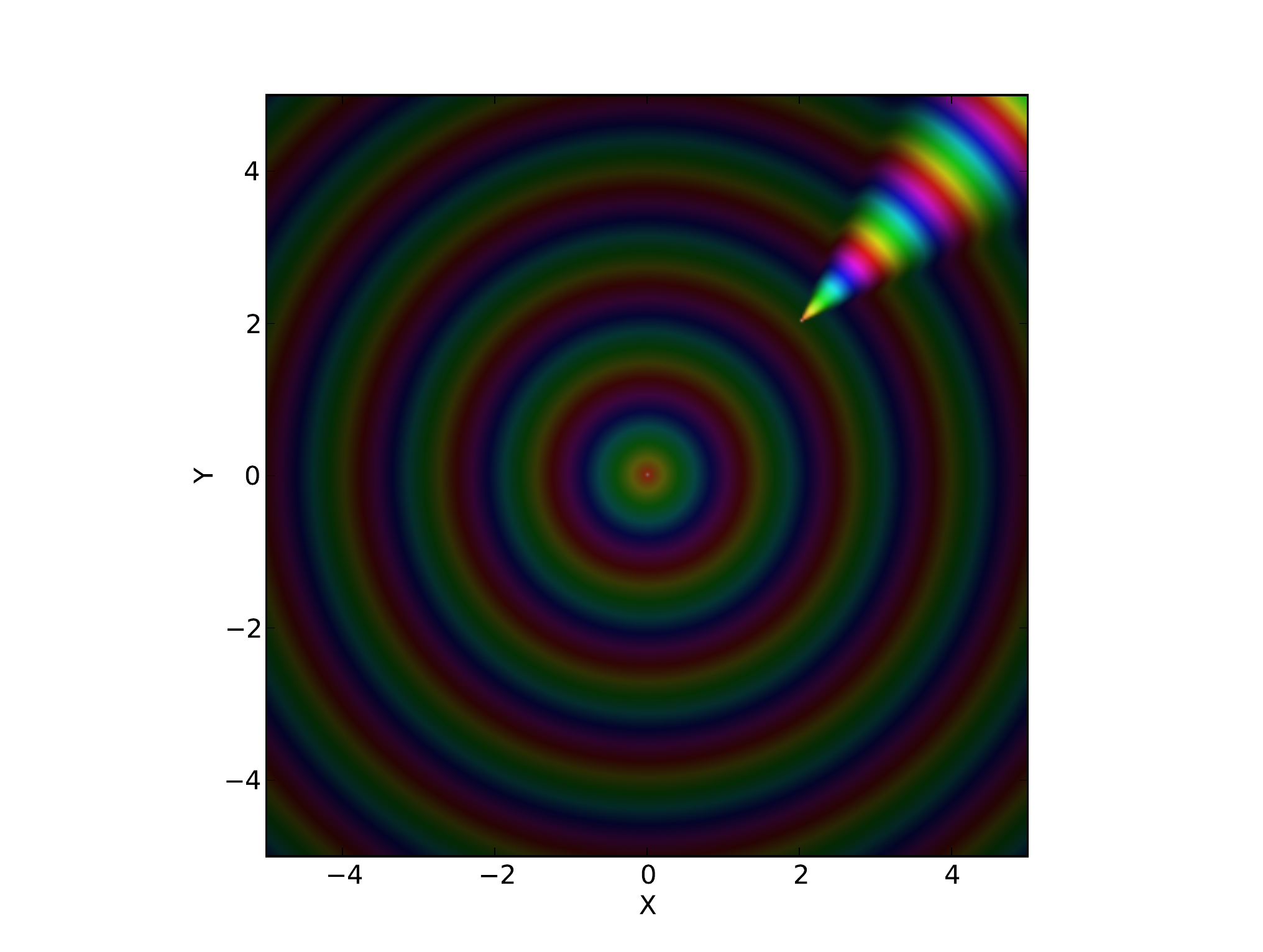}}}
 \put(0,5){(a)}
 \put(6.4,5){(b)}
\end{picture}}

 \caption{\label{fig:wf} $\alpha$-particle wave function (a) in the absence (b) in the presence of an obstacle.
In case (b), only the elastic-scattering part of the wave function is represented and
the wave function is assumed to have reached its asymptotic behaviour everywhere.
The phase of the wave function is represented by the colour hue while the modulus is represented by
the colour brightness \cite{thaller:00}.}
\end{figure}

Within these hypotheses, the stationary wave function of the total $\alpha$-particle + obstacle system
can be approximately described as a first-order Born-expanded coupled-channel wave function, which reads
\begin{equation}
\label{wf}
\Psi(\bm{R}, \bm{r})
\approx f^{(0)}_0(\bm{R}) \psi_0(\bm{r}) + f^{(1)}_0(\bm{R}) \psi_0(\bm{r}) + f^{(1)}_1(\bm{R}) \psi_1(\bm{r}).
\end{equation}
In this equation, the superscript refers to the order in the Born expansion
while the subscript refers to the excitation state of the obstacle.
The first term corresponds to the situation
where the $\alpha$ particle is unaffected by the presence of the obstacle;
its wave function is thus an outgoing spherical wave of kinetic energy $E_\alpha$,
\begin{equation}
\label{sph}
 f^{(0)}_0(\bm{R}) = \frac{e^{i k R}}{R}, \qquad k = \sqrt{2 m_\alpha E_\alpha/\hbar^2}.
\end{equation}
This wave function is represented in figure \ref{fig:wf}(a);
in this figure, the exponential phase factor is visible as coloured rings,
while the $1/R$ modulus, necessary for flux conservation, is visible as decreasing brightness.
The second term of \eqref{wf} is the first-order Born expansion of the elastic scattering;
the wave function $f^{(1)}_0(\bm{R})$ hence corresponds to the same kinetic energy $E_\alpha$
as the zero-order term.
Note that this term was not considered by Mott in \cite{mott:1929}.
The third term of \eqref{wf} corresponds to the excitation of the obstacle by the $\alpha$ particle;
hence, by energy conservation, the wave function $f^{(1)}_1(\bm{R})$ is characterized by an energy $E_\alpha'$
and a wave number $k'$ which are defined as
\begin{equation}
 E_\alpha' = E_\alpha + E_0-E_1 \equiv \frac{\hbar^2 k'^2}{2m_\alpha}.
\end{equation}

These first-order terms can be calculated with the help of the Green-function method and read
\begin{eqnarray}
 f_0^{(1)}(\bm{R}) & = & \frac{1}{4\pi} \int d\bm{R}' \frac{2 m_\alpha}{\hbar^2} V_{00}(\bm{R}'-\bm{a})
\frac{e^{ikR'}}{R'} \frac{e^{ik|\bm{R}-\bm{R}'|}}{|\bm{R}-\bm{R}'|}, \\
 f_1^{(1)}(\bm{R}) & = & \frac{1}{4\pi} \int d\bm{R}'\frac{2 m_\alpha}{\hbar^2} V_{10}(\bm{R}'-\bm{a})
\frac{e^{ikR'}}{R'} \frac{e^{ik'|\bm{R}-\bm{R}'|}}{|\bm{R}-\bm{R}'|}, \\
\end{eqnarray}
with
\begin{equation}
\label{Vj0}
 V_{j0}(\bm{R}) \equiv \int d\bm{r} \psi_j^*(\bm{r}) V(\bm{R}, \bm{r}) \psi_0(\bm{r}).
\end{equation}
The remarkable feature of these wave functions, as already stressed by Mott \cite{mott:1929},
is that they present an asymptotic behaviour strongly peaked along direction $\hat{\bm{a}}$.
To show that, we first assume that the energy excitation of the obstacle is negligible with respect
to the $\alpha$-particle kinetic energy, which implies $k'\approx k$.
Then, by defining the unit vector
\begin{equation}
 \hat{\bm{e}} \equiv \frac{\bm{R}-\bm{a}}{|\bm{R}-\bm{a}|}
\end{equation}
and the angle $\theta$ between $\hat{\bm{a}}$ and $\hat{\bm{e}}$,
\begin{equation}
 \theta=\arccos(\hat{\bm{a}} \cdot \hat{\bm{e}}),
\end{equation}
one gets for the asymptotic behaviour of the first-order wave functions
\begin{equation}
\label{fj1as}
 f_j^{(1)}(\bm{R}) \mathop{\approx}_{a, |\bm{R}-\bm{a}| \gg s}
\frac{e^{ik|\bm{R}-\bm{a}|}}{|\bm{R}-\bm{a}|} I_j(\theta).
\end{equation}
That is a spherical wave emitted in $\bm{a}$,
modulated by an angular-dependent function
\begin{equation}
\label{Ij}
I_j(\theta) = \frac{m_\alpha}{2\pi\hbar^2} \frac{e^{ika}}{a}
\int d\bm{R}' V_{j0}(\bm{R}') e^{i\bm{q}\cdot\bm{R}'},
\end{equation}
where we have defined the transferred momentum
\begin{equation}
 \bm{q} \equiv k(\hat{\bm{a}}-\hat{\bm{e}}),
\end{equation}
which is related to $\theta$ by
\begin{equation}
 q = 2k \sin\frac{\theta}{2}.
\end{equation}
To establish \eqref{fj1as} and \eqref{Ij},
we have used the fact that matrix elements \eqref{Vj0} only have a non negligible value
in a volume of size $s$ around the obstacle position $\bm{a}$.
Expression \eqref{Ij} contains the Fourier transform of matrix elements \eqref{Vj0},
a well-known result for the first-order Born approximation.
These matrix elements can be calculated explicitly for various assumptions on the obstacle,
in particular in the case of hydrogen-like wave functions \cite{mott:1929,taylor:72}.
Here, as we are rather interested in general properties of the scattering wave function \eqref{wf},
we simply assume that the matrix elements \eqref{Vj0} have a spherical Gaussian form factor of width $s$.
This allows us to plot a typical behaviour of the wave function in panel (b) of figure \ref{fig:wf}.
In this figure, only the spherical and peaked elastic-scattering waves are shown;
the inelastic-scattering wave would have a behaviour similar to the elastic-scattering one
but with a broader peak.

Let us now come to the delicate point of the wave-function normalization,
which is not discussed in \cite{mott:1929}.
Here, we fix this normalization by imposing that the total probability flux $F$ across
a sphere centred on the $\alpha$ emitter and large enough to include the obstacle
should be the same whether this obstacle is present or not.
This seems natural,
as the activity of the radioactive source should be independent of its surroundings,
but we shall show that it has important physical consequences.
Without obstacle, this probability flux takes the value
\begin{equation}
\label{Fwithout}
F_\mathrm{without}= 4\pi \frac{\hbar k}{m_\alpha} \equiv 4\pi v_\alpha,
\end{equation}
which only depends on the $\alpha$-particle velocity $v_\alpha$.
This derives from the probability current
\begin{equation}
 \bm{J}=\frac{1}{m_\alpha}\Re\left[f_0^{(0)*}(\bm{R})
\left(-i\hbar \nabla_{\bm{R}} \right) f_0^{(0)}(\bm{R})\right]
\end{equation}
corresponding to the spherical wave function \eqref{sph},
integrated on all angles on the large-radius sphere.
With an obstacle, the calculation is more complicated as a projection on the obstacle states has to be made
and the current calculation makes an interference term appear between the first two terms of \eqref{wf}.
However, these interference terms vanish when the current is integrated on a large enough sphere.
Hence, the total flux reads
\begin{equation}
 F = 4 \pi v_\alpha + 2\pi v_\alpha \int_0^\pi d\theta \sin \theta |I_0(\theta)|^2 
+ 2\pi v'_\alpha \int_0^\pi d\theta \sin\theta |I_1(\theta)|^2.
\end{equation}
Since the last two terms of this expression are clearly positive,
this flux is larger than \eqref{Fwithout}.
Hence, the coupled-channel wave function \eqref{wf} has to be multiplied by a factor $C$,
with a modulus smaller than 1, defined as
\begin{equation}
 |C|^2 = \left[1+\frac{1}{2} \int_0^\pi d\theta \sin \theta |I_0(\theta)|^2
+ \frac{1}{2} \frac{v'_\alpha}{v_\alpha} \int_0^\pi d\theta \sin \theta |I_1(\theta)|^2\right]^{-1},
\end{equation}
for the fluxes to be equal with and without obstacle.
The effect of this factor is clearly visible on figure \ref{fig:wf}:
the amplitude of the spherical wave is reduced by the presence of the obstacle.
One has thus
\begin{equation}
 F_{\mathrm{spherical}}=|C|^2 F_{\mathrm{without}}, \quad |C|^2<1,
\end{equation}
which is the key result of this part of the paper.

Following Mott \cite{mott:1929}, we now consider the scattering on two successive obstacles located in $\bm{a}$
and $\bm{b}$ [see figure \ref{fig:syst}(b)].
%
%
A second-order Born expansion is required and the coupled-channel wave function reads
\begin{eqnarray}
 \Psi(\bm{R}, \bm{r}_a, \bm{r}_b)
& = & \left[f^{(0)}_{00} (\bm{R}) + f^{(1)}_{00} (\bm{R}) + f^{(2)}_{00} (\bm{R})\right]
\psi_0(\bm{r}_a) \psi_0(\bm{r}_b) \nonumber \\
&& + f^{(1)}_{10} (\bm{R}) \psi_1(\bm{r}_a) \psi_0(\bm{r}_b)
+ f^{(1)}_{01} (\bm{R}) \psi_0(\bm{r}_a) \psi_1(\bm{r}_b)
+ f^{(2)}_{11} (\bm{R}) \psi_1(\bm{r}_a) \psi_1(\bm{r}_b).
\end{eqnarray}
The very important result by Mott is that the second-order wave function $f^{(2)}_{11}$,
corresponding to an excitation of both atoms, is significantly
different from zero only if $\bm{a}$ and $\bm{b}$ are aligned
(for atoms $\bm{a}$ and $\bm{b}$ to be excited, atom $\bm{b}$ has to lie in the narrow cone
generated by the presence of atom $\bm{a}$).
This explains the appearance of linear tracks from an initially symmetric spherical wave function.
Here, we complete this result by noting that, in case of an alignment,
this wave function should be multiplied by a factor of the order
of $|C|^4$ for the probability flux to be equal to the flux without obstacles.
We infer that for $N$ successive aligned obstacles, the wave function should be multiplied
by a factor $|C|^{2N}$.
This would have a spectacular consequence on the spherical-wave flux, which would read
\begin{equation}
\label{Natoms}
 F_\mathrm{spherical} \approx|C|^{2N} F_\mathrm{without}.
\end{equation}
That is the spherical wave would tend to zero if there is a large enough number of atoms
in the cloud chamber aligned with the $\alpha$ emitter.
We conjecture this might be a model for the phenomenon of wave-function reduction in quantum mechanics.

Now, in the case of a cloud chamber consisting of randomly-distributed atoms,
this mechanism would select the direction of the atoms best-aligned with the $\alpha$ emitter:
only for these atoms would the high-order component of the wave function not vanish.
The wave-function reduction would thus only occur if some atoms are aligned in a given direction;
that direction would be the measured linear track.
This model thus provides a deterministic explanation to the apparent randomness of quantum measurement.

\section{Possible theoretical flaws and experimental tests}

One should keep in mind that the above reasoning rests on simplifying hypotheses,
which should be tested carefully in future works.
First, the calculation is based on the Born approximation,
which is well-known to violate unitarity \cite{taylor:72}.
We do not expect this to be problematic here as the high energy of the $\alpha$ particle
probably implies that the Born approximation is very good;
nevertheless, convergence tests should be carried out.
Second, the calculation is based on a truncated coupled-channel approximation;
there again, convergence tests including more excited states should be made.
These two first possible problems could be avoided by directly solving the Schrödinger equation numerically
and compare the obtained result with the approximated one.
Finally, last but not least, our interpretation is based on a stationary state calculation,
whereas it has a temporal content in essence:
the wave function is {\em first} affected by the atom nearest to the source,
{\em then} by the second nearest atom and so on.
This stationary-state approach is an approximation of the full solution of the time-dependent
Schrödinger equation.
There again, solving this equation numerically for instance for an initial spherical wave packet
would provide very useful checks.

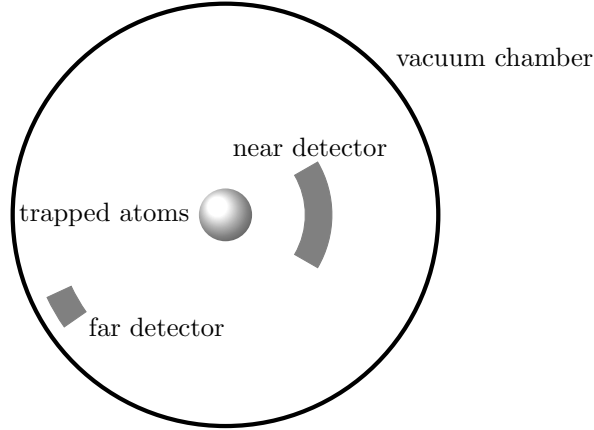
\begin{figure}
\centerline{
\begin{tikzpicture}[scale=0.7]
 \draw[ultra thick] (0,0) circle (4) (3,3) node[right] {vacuum chamber};
 \shade[ball color=white] (0,0) circle (0.5) (-0.5,0) node[left]
{trapped atoms};
 \filldraw[gray] (30:1.5) arc (30:-30:1.5) -- (-30:2) arc (-30:30:2) -- cycle
node[black,xshift=-1mm, yshift=2mm] {near detector};
 \filldraw[gray] (-145:3.2) arc (-145:-155:3.2) -- (-155:3.7) arc (-155:-145:3.7) -- cycle
node[black,right,xshift=2mm] {far detector};
\end{tikzpicture}}
 \caption{\label{fig:setup}Schematic experimental set-up for the detection by a far detector
of the reduction of the sphe\-rical waves emitted by $\alpha$-radioactive trapped atoms,
in the presence of a detector close to the emitter.}
\end{figure}

Let us now assume that the reduction of the spherical wave in the presence of an obstacle
shown in figure \ref{fig:wf} is correct
and briefly explore some set-ups that might be used to test it experimentally.
A first important aspect to consider is the practical implementation of a spherical-wave emission.
A solid-state $\alpha$ source might not be a good candidate as decoherence of the spherical wave
might occur in the source already.
An alternative option might be a mesoscopic ensemble of radioactive atoms (or ions),
trapped in an atomic trap as a low-density gas,
so as to limit the interactions between the emitted $\alpha$ particles and the other atoms of the source.
A second aspect is the nature of the obstacle leading to the spherical-wave reduction.
To display a substantial effect, the interaction between the $\alpha$ particle and the obstacle
should be strong.
According to \eqref{Natoms}, a possible way could be to align a large number of atoms,
which seems difficult to achieve in practice.
We might instead assume that the mechanism proposed above for the cloud chamber,
namely that the internal state of the apparatus determines the measurement result,
is actually valid for any kind of particle detector.
If this is true, replacing the obstacle by a high-efficiency detector covering a limited solid angle
would lead to an observable wave-function reduction in other directions.
This reduction could be observed by another detector placed at a larger distance from the $\alpha$-particle source,
as illustrated in figure \ref{fig:setup}.
A difficulty with such an experimental set-up might be to get an absolute flux measurement,
both in the presence and absence of the ``obstacle detector''.

Another option might be to use photons instead of $\alpha$-particles.
Electric dipole photons emitted by trapped polarized atoms are also characterized by a wide
(though not spherical) angular distribution.
Their interaction with an obstacle (say a single atom) in one particular direction might also
lead to a detectable flux reduction in other directions.
The advantage of such a set-up would be that the presence or absence of the obstacle could be simulated
by tuning or detuning, for instance by Zeeman effect,
an atomic transition of the obstacle atom with the energy of the electric-dipole photon,
hence strengthening or dimming their interaction.
On the other hand, absolute flux measurement and high-efficiency detection might be more difficult to achieve
with photons than with $\alpha$ particles.

\section{Speculative conclusions: faster-than-light or not?}

As a conclusion we propose an interpretation of quantum mechanics which is deterministic in essence.
The outcome of a decoherent process like a measurement can in principle be predicted from the
knowledge of the microscopic state of the environment or macroscopic apparatus.
This state being inaccessible in most practical situations,
the outcomes of decoherence processes generally seem random.
Nevertheless this randomness now appears as much less fundamental and unavoidable than
in standard quantum mechanics presentations.
We have first shown, without making any attempt to describe microscopic states explicitly,
that this interpretation leads to a violation of Bell's inequalities,
despite the fact that this microscopic state can be seen as a hidden variable.
Next we have studied explicitly a simplified model of measurement apparatus,
the cloud chamber,
and obtained results that support our interpretation:
the wave-function collapse leading to the measurement outcome
is determined by the positions of the atoms in the cloud chamber.
Let us stress that this interpretation, to put it on Mott's words \cite{mott:1929},
is based on ``wave mechanics unaided'':
no further ingredient (pilot wave, many worlds, free will\dots) than wave functions is required.

Let us next notice that both types of quantum states considered above are highly non local.
The EPR intricated Bell state \eqref{singlet} implies
a perfect and instantaneous correlation of the measured states of particles 1 and 2,
despite the fact that these states are not known in advance.
Similarly the $\alpha$-particle spherical state \eqref{sph} implies
that the detection of the particle in one direction immediately prevents the particle from being detected
in other directions.
These states themselves do not violate special relativity and causality
as they can be explained by their unique spatio-temporal origin
either at the creation of the pair or at the emission of the $\alpha$-particle.
When submitted to a measurement process
their non locality is revealed and an action at a distance is necessary
to explain the perfect correlations between space-separated events.
In usual interpretations of quantum mechanics,
causality is however preserved by the random character of the measurement results \cite{eberhard:78},
as already mentioned in the introduction.

In contrast, in the present interpretation this random character is replaced by a deterministic interpretation.
For the spherical wave, the presence of the obstacle in a given direction immediately leads to a
reduction of the spherical wave in all other directions,
even if the distance between the emitter and the obstacle can be made very large, at least in principle.
Hence by controlling the presence of the obstacle at a given position
one could immediately transfer information to remote space-separated regions,
as proposed on a small scale in figure \ref{fig:setup}.
For the EPR pair,
the state of apparatus 1 formally described by the value of $\Lambda_1$
determines the result of the measurement performed on particle 1.
To control $\Lambda_1$ in practice,
one could for instance think of a Stern-Gerlach apparatus with the screen replaced by a cloud chamber
or of a cold or mesoscopic polariser.
Particles 1 and 2 being fully intricated,
the value of $\Lambda_1$ also instantaneously determines the state of particle 2,
even though apparatus 1 and particle 2 can be separated by very large distances (or by very long optical fibers).
The present interpretation thus suggests
that quantum non locality actually allows faster-than-light information transfer.

This raises of course many paradoxes of special relativity,
which might be considered as a very strong ``no-go'' argument for our interpretation.
We rather speculate that our approach might finally settle the long-standing conflict
between special relativity and quantum mechanics,
Einstein's two warring daughters.
This conflict was made explicit by Einstein himself in the EPR paper,
then brought to a climax by Bell and his inequalities,
to finally burst into a triumphant victory of quantum mechanics in Aspect's famous experiments.
We believe the quantum world to be fully deterministic and non local
and hope this work will help testing this very strong hypothesis thoroughly.
As for special relativity paradoxes,
we adopt a very ``engineering-like'' approach
by suggesting to first build an instantaneous-information-transfer machine and, if it works,
to deal with the paradoxes it creates afterwards!

\section*{Acknowledgements}
JMS acknowledges very interesting discussions with several colleagues at different stages of this work,
in particular with N.~J.\ Cerf, D.\ Baye, J.\ Barbour, P.\ Gaspard, C.\ Semay and P.\ Capel.

\bibliographystyle{unsrt}
\bibliography{../../Biblio/others.bib}
\end{document}